%% file: IJRNC2024.tex
\RequirePackage{etex}
\providecommand{\reserveinserts}[1]{}
\let\NJDbibliographystyle\bibliographystyle
\def\bibliographystyle#1{\NJDbibliographystyle{WileyNJD-AMS-unsrt}}
\documentclass[AMS, Times1COL]{WileyNJDv5} 

\articletype{Research Article}%

\journal{International Journal of Robust and Nonlinear Control}
\volume{35}
\copyyear{2025}
\startpage{1}

\input{TeX/Preamble.tex}
\usepackage{comment}

\makeatletter
\def\NAT@sort{\@ne}\def\NAT@cmprs{\@ne}
\def\oddhead@titlepage@info{\hbox to \textwidth{\fontsize{8}{10}\selectfont
	This paper has been published in \emph{International Journal of Robust and Nonlinear Control}, 35(18):7621--7632, 2025.
	\href{https://doi.org/10.1002/rnc.70083}{doi:10.1002/rnc.70083}\hfill}}%
\def\evenhead@titlepage@info{\oddhead@titlepage@info}%
\def\oddfoot@titlepage@info{}%
\def\evenfoot@titlepage@info{}%
\makeatother

\begin{document}

\title{Enhanced Sampled-Data Model Predictive Control via Nonlinear Lifting}

\author[1]{Nuthasith Gerdpratoom}

\author[2]{Fumiya Matsuzaki}

\author[3]{Yutaka Yamamoto}

\author[1]{Kaoru Yamamoto}

\authormark{Gerdpratoom \textsc{et al.}}
\titlemark{Enhanced SD-MPC via nonlinear lifting}

\address[1]{\orgdiv{Graduate School and Faculty of Information Science and Electrical Engineering}, \orgname{Kyushu University}, \orgaddress{\state{Fukuoka}, \country{Japan}}}

\address[2]{\orgdiv{Joint Graduate School of Mathematics for Innovation}, \orgname{Kyushu University}, \orgaddress{\state{Fukuoka}, \country{Japan}}}

\address[3]{\orgdiv{Graduate School of Informatics}, \orgname{Kyoto University}, \orgaddress{\state{Kyoto}, \country{Japan}}}

\corres{Kaoru Yamamoto (yamamoto.kaoru.481@m.kyushu-u.ac.jp)}

\fundingInfo{JSPS KAKENHI Grant Number JP24K07546.}

\abstract[Abstract]{%
	This paper introduces a novel nonlinear model predictive control (NMPC) framework that incorporates a lifting technique to enhance control performance for nonlinear systems.
	While the lifting technique has been widely used in linear systems to capture intersample behavior, their application to nonlinear systems remains unexplored.
	We address this gap by formulating an NMPC scheme that combines fast-sample/fast-hold approximations and numerical methods to approximate system dynamics and cost functions. The proposed approach is validated through two case studies: the Van der Pol oscillator and the inverted pendulum on a cart.
	The simulation results demonstrate that the lifted NMPC outperforms conventional NMPC in terms of reduced settling time and improved control accuracy.
	These findings underscore the potential of the lifting-based NMPC for efficient control of nonlinear systems, offering a practical solution for real-time applications.
}


\jnlcitation{\cname{%
\author{Gerdpratoom N},
\author{Matsuzaki F},
\author{Yamamoto Y}, and
\author{Yamamoto K}}.
\ctitle{Enhanced Sampled-Data Model Predictive Control via Nonlinear Lifting.} \cjournal{\it Int J Robust Nonlinear Control.} \cvol{2025;35(18):7621--7632. doi:10.1002/rnc.70083}.}

\maketitle

\renewcommand\thefootnote{}
\footnotetext{\textbf{Abbreviations:} NMPC, nonlinear model predictive control; FSFH, fast-sample/fast-hold.}

\renewcommand\thefootnote{\arabic{footnote}}
\setcounter{footnote}{0}


	\section{Introduction}\label{sec:introduction}%
		\input{TeX/Text/Introduction.tex}

	\section{Overview of Lifting Techniques}\label{sec:lifting}%
		\input{TeX/Text/Lifting.tex}

	\section{Nonlinear Model Predictive Control via Lifting}\label{sec:NMPC}%
		\input{TeX/Text/NMPC.tex}

	\section{Numerical Examples}\label{sec:examples}
		\input{TeX/Text/Examples.tex}

	\section{Conclusion}\label{sec:conclusion}
			\input{TeX/Text/Conclusion.tex}

	\bibliography{TeX/Liftingrefs.bib}



\end{document}

%% file: TeX/Text/Introduction.tex
The lifting technique \cite{bamieh1991lifting,yamamoto1994function} has been instrumental in advancing the modern theory of linear sampled-data control systems \cite{yamamoto2013control,yamamoto2014optimal,yamamoto2018digital}.
The fundamental challenge in sampled-data control lies in handling two different time sets: one discrete and the other continuous.
Traditional approaches to sampled-data synthesis either design an analogue controller followed by discretization or discretize the continuous-time plant before designing a digital controller.
Both approaches are bound to introduce approximations through discretization and have inherent limitations: the former does not explicitly account for the length of the sampling period in controller design, whereas the latter only addresses performance at sampling instants.
In contrast, the lifting technique can be regarded as an {\em exact} discretization method.
It transforms intersample continuous-time signals into elements of a function space, allowing continuous-time behavior to be represented as a discrete-time system with infinite-dimensional input and output vectors.
This enables the representation of sampled-data systems as discrete-time, time-invariant systems while maintaining intersample behavior.
An important consequence of introducing the lifting technique is that various control strategies, such as linear quadratic regulators and \(H^\infty/H^2\) control, can be well incorporated while fully accounting for intersample dynamics \cite{bamieh1992general,chen1995optimal}.
Since its introduction, the lifting technique has been successfully applied to achieve ripple-free tracking \cite{hara1996modern}, high-performance signal processing \cite{yamamoto2012signal}, stability analysis involving intersample dynamics \cite{smith1996stability, yamamoto2017renewed}, and more recently, to enable signal tracking and disturbance rejection beyond the Nyquist frequency when combined with multirate techniques \cite{yamamoto2017simultaneous,yamamoto2022hypertracking,yamamoto2016tracking}.

A natural question then arises: {\em How should the nonlinear counterpart be defined?}
Until recently, however, this fundamental issue has remained largely unaddressed.
A key challenge in extending lifting-based techniques to nonlinear systems lies in how the input affects the output.
In linear systems, this relationship can be captured by the fictitious direct feedthrough term, which represents the instantaneous effect of the input on the output.
This allows the system to be lifted by lifting only the input and output signals, without explicitly involving the state.
However, in nonlinear systems, such a separation is generally not possible---the output depends on the input through the evolution of the state.

Recently, the authors have addressed this issue by further lifting the state trajectory and proposed a new lifting technique for nonlinear systems \cite{yamamoto2023nonlinear}.
However, this framework requires solving a nonlinear ordinary differential equation, which typically lacks a closed-form solution.
Thus, incorporating numerical methods into the nonlinear lifting framework becomes essential for controller design.

In this study, we first utilize the fast-sample/fast-hold (FSFH) approximation \cite{chen1995optimal} alongside numerical integration methods to approximate solutions.
We then propose a nonlinear model predictive control (NMPC) framework incorporating this approximated lifted nonlinear system.
NMPC is a powerful tool for addressing nonlinear control problems by iteratively solving a finite-horizon optimal control problem with state and input constraints at each sampling instant, using updated measurements, thereby providing robustness to uncertainties.
Due to its flexibility in imposing control objectives and handling constraints directly in the controller design process, numerous active studies on both linear MPC and NMPC have been conducted across diverse areas in recent years \cite{schwenzer2021review, yu2021model}.
One of the NMPC's advantages is its direct handling of nonlinear dynamics at the cost of increased computational effort \cite{biegler2021perspective}.
Its effectiveness has been demonstrated in applications such as vehicle steering \cite{stano2023enhanced} and quadcopter trajectory tracking \cite{sun2022comparative}, where NMPC outperforms traditional or flatness-based controllers.
It has also been successfully applied to multirobot flocking control \cite{nag2022flock, nag2024distributed}, including considerations of real-time implementation \cite{gerdpratoom2025decentralized}.

However, standard NMPC formulations solve an optimization problem based on the costs evaluated at each sampling instant only.
This approach not only neglects the incorporation of intersample behavior into the controller design but also fails to ensure that constraints are satisfied between sampling instants \cite{kellett2023introduction}.
Our proposed NMPC effectively overcomes these limitations, leveraging the lifting technique.

Numerical simulations demonstrate that our method enhances performance in nonlinear control by effectively suppressing unwanted overshoot and undershoot.
In particular, the introduction of a multirate lifted NMPC framework ensures that control specifications are met even at slow sampling rates.
Similar improvements were attempted in \cite{oishi2021optimal} using a numerical approach to optimal control in nonlinear sampled-data systems, however, without a lifting framework.

The main contributions of this paper are summarized as follows:
\begin{itemize}
	\item The lifting technique is extended to nonlinear systems by numerically approximating the lifted dynamics via FSFH and integration methods, enabling practical controller design.
	\item A novel NMPC framework is developed that incorporates the approximated lifted nonlinear system, explicitly handling intersample dynamics during control synthesis.
	\item A multirate version of the lifted NMPC is introduced to enhance performance under slow sampling by enforcing constraints and objectives on a finer time scale.
	\item The effectiveness and real-time feasibility of the proposed methods are validated through numerical simulations on standard nonlinear benchmarks, demonstrating improved transient behavior and robustness compared to conventional NMPC, with solve times that remain acceptable for real-time implementation.
\end{itemize}

The outline of the paper is as follows: \cref{sec:lifting} introduces the concept of the lifting technique, beginning with the linear formulation and extending to the nonlinear version previously proposed by the authors \cite{yamamoto2023nonlinear}.
In \cref{sec:NMPC}, we formulate a general control problem within the NMPC framework, incorporating the lifting formulation.
\cref{sec:examples} demonstrates the effectiveness of the proposed method through two standard nonlinear control problems: control of a Van der Pol oscillator and swing-up control of an inverted pendulum on a cart.
Finally, \cref{sec:conclusion} concludes the paper.

%% file: TeX/Text/Lifting.tex
In this section, we first review the concept of linear system lifting, followed by the introduction of the nonlinear lifting technique.
\subsection{Linear System Lifting}
\begin{figure}
		\centering
		\begin{overpic}[scale = 0.8]{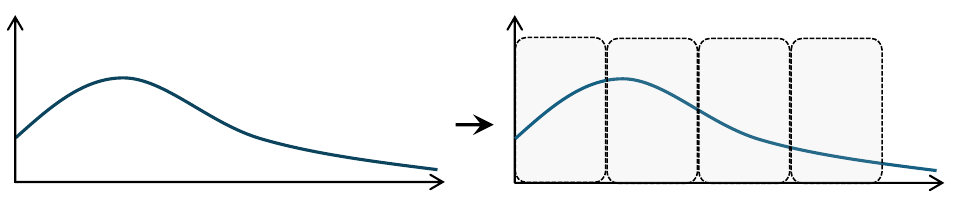}
			\put(9,14){\(\psi(t)\)}
			\put(46,-1){\(t\)}
			\put(54.5,14){\(\psi[0](\theta)\)}
			\put(64,14){\(\psi[1](\theta)\)}
			\put(73.5,14){\(\psi[2](\theta)\)}
			\put(83,14){\(\psi[3](\theta)\)}
			\put(98,-1){\(k\)}
			\put(52,-1){0}
			\put(62,-1){\(T\)}
			\put(71.5,-1){\(2T\)}
			\put(81,-1){\(3T\)}
			\put(90.5,-1){\(4T\)}
		\end{overpic}
		\caption{Concept of lifting.}
		\label{fig:lifting}
	\end{figure}

Consider the following linear time-invariant system:
	\begin{equation}
		\begin{aligned} \label{eq:linsys}
			\diff*{}x(t) &= Ax(t) + Bu(t)\\
			y(t) &= Cx(t) ,
		\end{aligned}
	\end{equation}
where \(x, u, y\) represent the state, input, and output, respectively, and \(A,B,C\) are constant matrices of suitable dimensions.
For an arbitrary \(T>0\) (which may later serve as the sampling period), define the {\em lifting\/} operator \(\lift\) as:
	\begin{equation}\label{eq:lifting}
		\lift: \psi \mapsto \{\psi[k](\cdot)\}_{k=0}^{\infty},\quad \psi[k](\theta) \coloneqq \psi(kT+\theta),
	\end{equation}
for \(0\leq \theta < T\), where \(\psi\) belongs to an appropriately defined function space on \([0, \infty)\), such as \(L^2\), \(L^2_{\rm loc}\), etc.
In other words, lifting maps a continuous-time signal to a discrete-time signal with function components (see \cref{fig:lifting}).
By lifting the input and output \cite{bamieh1991lifting}, the system \eqref{eq:linsys} can be expressed as:
	\begin{equation}
		\begin{aligned} \label{eq:lifted1}
			x[k+1] & = e^{AT}x[k] + \int_{0}^{T} e^{A(T-\tau)}Bu[k](\tau){\rm d}\tau \\
			y[k](\theta) & = Ce^{A\theta}x[k] + \int_{0}^{\theta} Ce^{A(\theta-\tau)}Bu[k](\tau){\rm d}\tau,
		\end{aligned}
	\end{equation}
where \(x[k]\) denotes the state \(x(t)\) at time \(kT\).
Observe that these expressions take the form
	\begin{equation}
		\begin{aligned} \label{eq:linearlifting}
			x[k+1] & = {\mathcal A} x[k] + {\mathcal B} u[k] \\
			y[k] & = {\mathcal C} x[k] + {\mathcal D} u[k],
		\end{aligned}
	\end{equation}
where the operators \(\mathcal{A},\mathcal{B},\mathcal{C},\mathcal{D}\) are independent of \(k\).
In other words, the continuous-time system \eqref{eq:linsys} has now been converted into a {\em time-invariant\/} discrete-time model with inputs and outputs in suitable function spaces.
However, note that the direct feedthrough term \(\mathcal{D}\) appears in \eqref{eq:linearlifting}, although it is absent in the original continuous-time system.
This direct influence of the input on the output, bypassing state transitions, is possible only in linear systems.
As a result, the state itself must also be lifted in the case of nonlinear systems.

\subsection{Nonlinear System Lifting}\label{sec:nonlinearlifting}
This section reviews the nonlinear lifting framework originally proposed in \cite{yamamoto2023nonlinear}.
Consider the following continuous-time nonlinear system:
	\begin{equation}
	\begin{aligned} \label{eq:nonlinearsystem}
		\diff*{}x(t) & = f(x(t), u(t)) \\
					y(t) & = h(x(t)) \\
					x(0) & = x_{0},
	\end{aligned}
	\end{equation}
where \(x(t) \in \R^{n}\).
We assume suitable conditions on \(f\) to ensure the existence and uniqueness of the solution, such as the Lipschitz condition with a global constant to prevent a finite escape time.
We also impose appropriate regularity conditions on \(h\).
The solution \(x(t)\) of this nonlinear differential equation is then denoted by \(\Phi(t, x_{0}, u)\).

We then define the following nonlinear discrete-time system by lifting the input, output, and state for \(T>0\):
	\begin{equation}
	\begin{aligned} \label{eq:liftedsys}
		x[k+1](\theta)	& = \Phi\left((k+1)T+\theta, x[k](T), u[k+1](\cdot)\right) \\
		y[k](\theta)	& = h(x[k](\theta)).
	\end{aligned}
	\end{equation}
Here, \(u[k+1](\cdot)\) indicates that \(x[k+1](\theta)\) depends on \(u[k+1](\tau),\) \(0\leq \tau \leq \theta\).
We refer to this system as the \emph{lifted system} of system (\ref{eq:nonlinearsystem}).
\begin{remark}
There are certain technical subtleties, such as the well-definedness of the output at \(t=kT+\theta\), which are discussed further in the literature \cite{yamamoto2023nonlinear}.
\end{remark}

\begin{figure}[tb]
	\centering
	\includegraphics{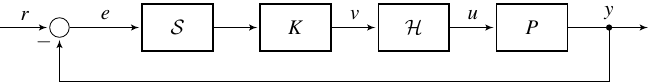}
	\caption{Unity feedback nonlinear sampled-data system.}
	\label{fig:sdunityfb}
\end{figure}

Note, however, that this system does not satisfy the strict causality condition, as \(u[k+1]\) appears on the right-hand side.
To address this lack of strict causality, we construct a feedback loop consisting of a strictly proper controller.
Consider the unity feedback sampled-data system in \cref{fig:sdunityfb}, where \(P\) is a nonlinear continuous-time plant described by \eqref{eq:nonlinearsystem},
and \(K\) is a discrete-time strictly proper controller:
	\begin{equation}
	\begin{aligned} \label{eq:controller}
		z[k+1]	&= \phi(z[k],e[k](0)) \\
		v[k]	&= \psi(z[k]),
	\end{aligned}
	\end{equation}
where \(z[k]\in \R^{n_c}\), \(e[k]\in\R^{m_c}\), and \(v[k]\in\R^{p_c}\).
Here, we assume that the error \(e(t)\) is uniformly sampled every \(T\) second by a sampler \(\mathcal{S}\), and the controller output \(v\in\R^{p_c}\) is held by a hold device \(\mathcal{H}\), producing a continuous-time control input \(u\in\R^m\) as
	\begin{equation}\label{eq:holdfun}
			u[k](\theta) = H(v[k]), \quad k=0,1,\dots,
	\end{equation}
where \(H(\cdot)\) is an appropriate hold function.

The lifted plant with period \(T\) is then given by \eqref{eq:liftedsys}.
Considering the delay induced by the strictly proper controller, we obtain the following strictly causal closed-loop system:
	\begin{equation}
	\begin{aligned} \label{eq:fbsys}
		x[k+1](\theta)	& = \Phi\left((k+1)T+\theta, x[k](T), H(v[k])\right) \\
		z[k+1]			& = \phi(z[k],e[k](0))\\
		v[k]			& = \psi(z[k])\\
		e[k](\theta)	& = r[k](\theta) - y[k](\theta).
	\end{aligned}
	\end{equation}

In the sequel, we focus on designing an MPC controller for \(K\).

%% file: TeX/Text/NMPC.tex
Consider again the nonlinear continuous-time plant \eqref{eq:nonlinearsystem} and formulate an MPC problem to design a discrete-time controller \(K\) for the unity sampled-data feedback system in \cref{fig:sdunityfb}, with a sampling period \(T\).

One may first approximate the continuous-time plant \eqref{eq:nonlinearsystem} by 
	\begin{equation}\label{eq:approxplant}
		x_{\rm d}[k+1] = F(x_{\rm d}[k],u_{\rm d}[k]), \quad x_{\rm d}[0] = x_{\rm{d}0},
	\end{equation}
using a discretization scheme, and then solve the following optimal control problem for a given prediction horizon \(N\in\N\cup \{\infty\}\) at each sampling instant \(t=nT\), \(n\in \Z\):
	\begin{equation} \label{eq:fin_mpc}
		\begin{aligned}
			&\minimise_{u_{\rm d}[k]\in \R^m , \, k =0,\dots, N-1} && \sum_{k=0}^{N-1} \ell(x_{\rm d}[k], u_{\rm d}[k]) + \varphi(x_{\rm d}[N]) \\
			&\hspace{8pt}\stt &&	\left. x_{\rm d}[0] = x(nT) \right.\\
			&		&&	\left. \begin{aligned}
									&x_{\rm d}[k+1] = F(x_{\rm d}[k], u_{\rm d}[k]) \\
									&x_{\rm d}[k] \in \mathcal{X}\\ 
									&u_{\rm d}[k] \in \mathcal{U}\\
								\end{aligned}\right\} \ k = 0,\dots,N-1\\
			&		&&	\left.  x_{\rm d}[N] \in \mathcal{X}_f, \right.
		\end{aligned}
	\end{equation}
where \(\ell:\R^n\times\R^m \to \R\) and \(\varphi:\R^n \to \R\) represent the stage and terminal costs, respectively.
This optimization is iteratively performed every \(T\) second with \(u_{\rm d}[0]\) being applied to the system at each iteration.

However, this approach only enforces the constraints and minimizes the cost at each sampling instant.
To incorporate the intersample dynamics into the design, we introduce the nonlinear lifting technique from \cref{sec:nonlinearlifting} into the NMPC framework.

\subsection{Enhanced NMPC via Lifting}
Now, we present an ideal NMPC formulation that incorporates nonlinear lifting.
Instead of \eqref{eq:fin_mpc}, we address the following optimal control problem at each sampling instant \(t=nT\), \(n\in \Z\):
	\begin{equation} \label{eq:liftedmpc}
		\begin{aligned}
			&\minimise_{v[k]\in\R^{p_c}, \, k =0,\dots, N-1} && \sum_{k=0}^{N-1}\int_0^T \ell(x[k](\theta), u[k](\theta)){\rm d}\theta + \varphi(x[N](0))  \\
			&\hspace{8pt}\stt	&&	\left. x[0](\theta) = \Phi(\theta,x(nT),v[0]) \right.\\
			&		&&	\left. \begin{aligned}
									&x[k+1](\theta) = \Phi((k+1)T+\theta,x[k](T),H(v[k]))\\
									&x[k](\theta) \in \mathcal{X}\\ 
									&u[k](\theta) \in \mathcal{U}\\
								\end{aligned}\right\} \ k = 0,\dots,N-1\\
			&		&&	\left.  x[N](0) \in \mathcal{X}_f, \right.
		\end{aligned}
	\end{equation}
where \(0\leq\theta<T\) and \(u[k](\theta) = H(v[k])\) as in \eqref{eq:holdfun}.

As is well recognized \cite{nesic2015nonlinear,nesic2007sampled}, an explicit solution of system \eqref{eq:nonlinearsystem}, \(\Phi(t,x_0,u)\), is generally unobtainable except in rare cases.
To address this challenge, we use an FSFH approximation by subdividing the sampling interval \([0,T]\) at each sampling period into segments \([0,T/N'),[T/N',2T/N'),\dots,[T-T/N',T)\).
We then apply a suitable numerical method for ordinary differential equations, such as the Runge--Kutta methods, to approximate \(x[k](T/N'),x[k](2T/N'),\dots,x[k](T)\) for \eqref{eq:nonlinearsystem}, thereby obtaining an approximation of \(\Phi(\cdot,x_0,u)\).
It is important to note that we do not require the actual sampling of the trajectory; instead, this is merely an artificial approximation of the lifted system.
To compute the integration in the cost function of \eqref{eq:liftedmpc}, a numerical integration method such as Simpson's rule can be applied.

An implementation of the lifted NMPC is described in \cref{alg:liftedmpc}, where we assume a multirate control scheme with an upsampling factor \(M\) and a zero-order hold.
In this case, the control input \(u[k](\theta)\) is defined as
	\begin{equation}\label{eq:multirateinput}
		u[k](\theta)	= H(v[k]) 
						=	\begin{cases}
								v_1[k] & (0\leq \theta < T/M)\\
								v_2[k] & (T/M\leq \theta < 2T/M)\\
								\hfil \vdots &\\
								v_M[k] & ((M-1)T/M\leq \theta < T),\\
							\end{cases}
	\end{equation}
where \(u[k](\theta) \in \R^m\) and \(v[k] \eqqcolon \begin{bmatrix} v_1[k]^\top, \dots, v_M[k]^\top\end{bmatrix}^\top \in\R^{mM}.\)
Note that \(M=1\) corresponds to a standard single-rate control using a zero-order hold, where the control input remains constant throughout the sampling interval.


\begin{algorithm}
\caption{Enhanced NMPC via lifting.}
\label{alg:liftedmpc}
\begin{algorithmic}[1]
\Require Prediction horizon $N$; subdivision factor $N'$; upsampling factor $M$; sampling period $T$.
	\For{$n = 0, 1, 2, \dots$}
		\State Measure or estimate the current state \(x(nT)\) and define \(x[0](0) = x(nT)\) 
		\State Solve the optimal control problem \eqref{eq:liftedmpc} to obtain the optimal control sequence \(v[k]\in\R^{mM}, k = 0,\dots, N-1\)
		\State Actuate \(u[0](\theta)\), \(0\leq \theta < T\), where \(u[0](\theta)\) is given by \eqref{eq:multirateinput}
	\EndFor
\end{algorithmic}
\end{algorithm}

%% file: TeX/Text/Examples.tex
In this section, we focus on the following quadratic stage and terminal costs in \eqref{eq:fin_mpc} and \eqref{eq:liftedmpc}:
	\begin{equation}
		\ell(x,u) = x^\top Q x + u^\top R u, \quad \varphi(x) = x^\top Q' x,
	\end{equation}
where \(Q \succeq 0, R \succ 0, Q' \succeq 0\) are weight matrices.
We also assume the zero-order hold so that the control input remains constant throughout the sampling interval.


We now present two examples: the steering control of the Van der Pol oscillator and the swing-up control of an inverted pendulum.
In both examples, we use the fourth-order Runge--Kutta method to approximate \(\Phi(\cdot,x_0,u)\) and Simpson's rule to integrate the cost function in \eqref{eq:liftedmpc}.

\subsection{Steering Control of a Van der Pol Oscillator}
\begin{figure}[ht]
		\centering
		\begin{subfigure}{0.48\linewidth}
			\centering
			\includegraphics[width=\linewidth]{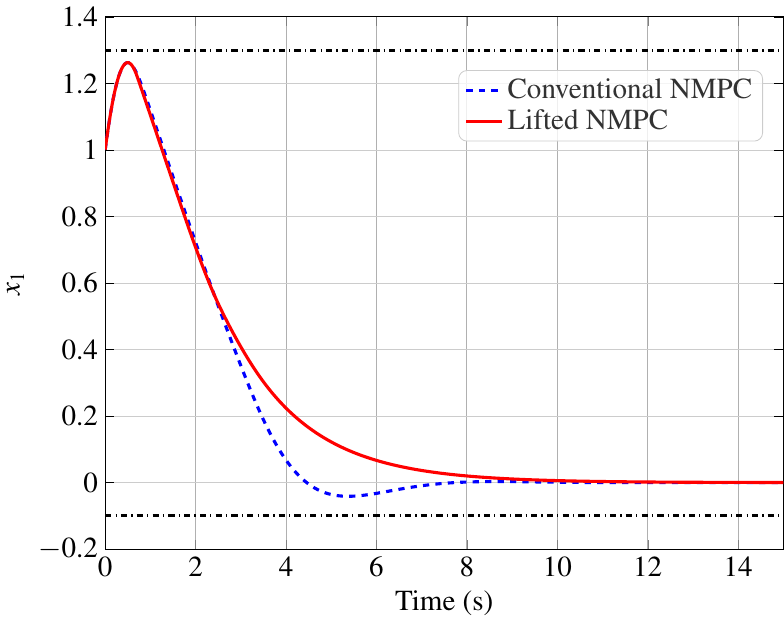}%
		\end{subfigure}
		\hfill
		\begin{subfigure}{0.48\linewidth}
			\centering
			\includegraphics[width=\linewidth]{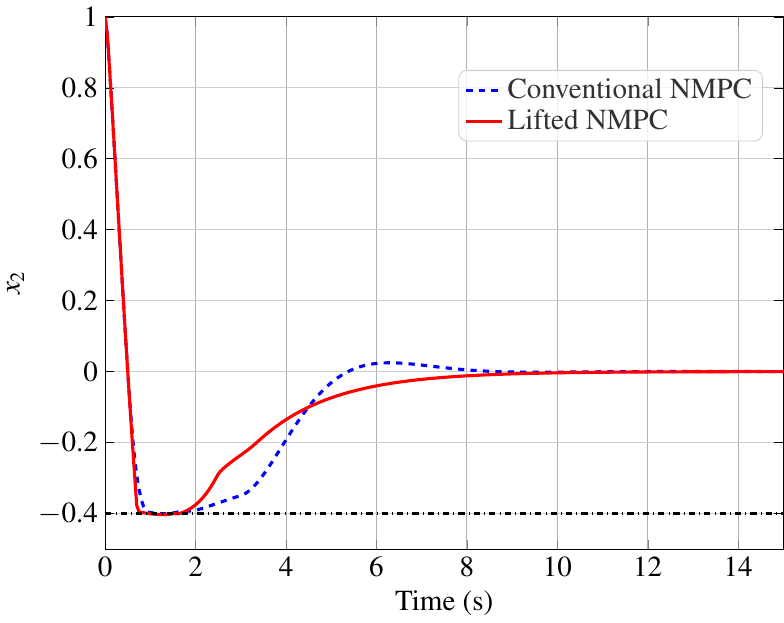}%
		\end{subfigure}

		\vspace{0.5cm} 

		\begin{subfigure}{0.48\linewidth}
			\centering
			\includegraphics[width=\linewidth]{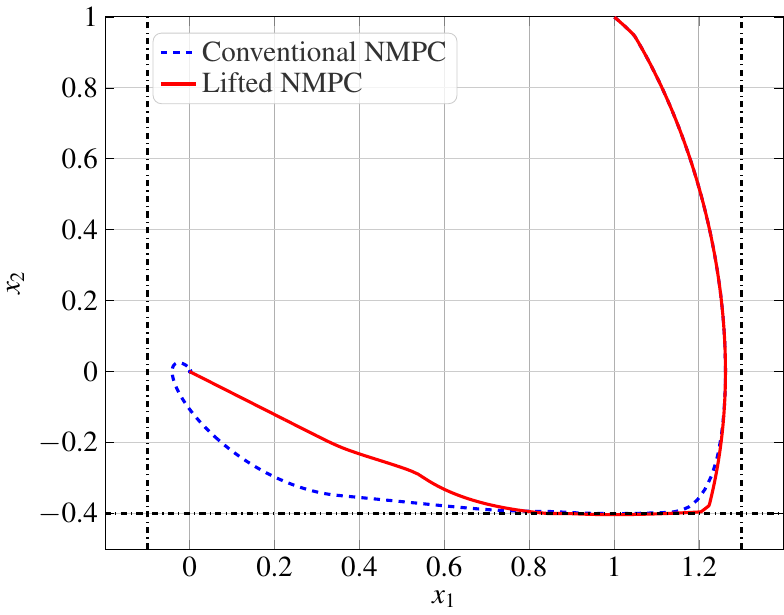}%
		\end{subfigure}
		\hfill
		\begin{subfigure}{0.48\linewidth}
			\centering
			\includegraphics[width=\linewidth]{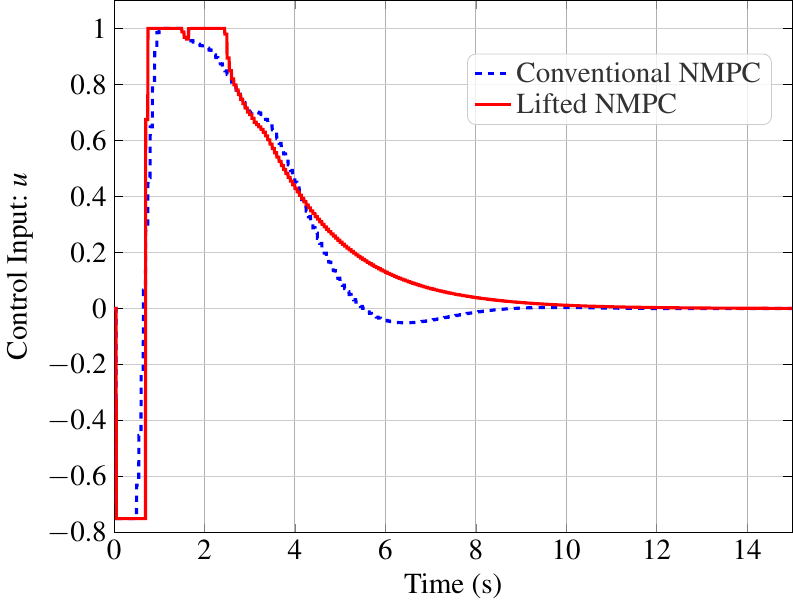}%
		\end{subfigure}

		\caption{Comparison of the Van der Pol oscillator's behavior under conventional NMPC (blue dashed) and lifted NMPC (red solid).
		Time series trajectories of \(x_1\) (top-left) and \(x_2\) (top-right) are shown, along with the system's phase portrait (bottom-left) and the control input \(u\) (bottom-right). Black dashed-dotted lines represent the state constraints.}
		\label{fig:combined_vdp}
	\end{figure}

Consider the problem of steering the Van der Pol oscillator:
	\begin{equation} \label{eq:vdp}
		\dot{x}(t) =
		\begin{bmatrix}
		\dot{x}_{1}(t) \\
		\dot{x}_{2}(t)
		\end{bmatrix}
		=
		\begin{bmatrix}
		x_2(t) \\
		-\mu(x_1^2(t) - 1)x_2(t) - x_1(t) + u(t)
		\end{bmatrix},
	\end{equation}
where \(x\coloneqq [x_1,x_2]^\top\) represents the state variables, \(\mu > 0\) is a damping parameter (set to \(\mu = 1\) for the simulation), and \(u\) is the control input.
In this example, we consider single-rate control with a zero-order hold both for the conventional discrete-time NMPC and the proposed enhanced NMPC via lifting.
Specifically, the control input for the former is given by \(u(t) = u_{\rm d}[k]\), and for the latter by \(u(t) = v[k]\) (i.e., \(M=1\) in \eqref{eq:multirateinput}) for \(kT\leq t < (k+1)T\), \(k=0,1,\dots\). 
The inputs \(u_{\rm d}[k]\) and \(v[k]\) are computed by solving \eqref{eq:fin_mpc} and \eqref{eq:liftedmpc}, respectively, using the following parameters:
	\begin{itemize}
		\item Sampling period: \(T=0.05\) (s)
		\item Prediction horizon: \(N=5\)
		\item Cost function weights:
			\(
				Q = \diag(4,1),
				\quad
				R = 1,
				\quad
				Q' = 2Q
			\)
		\item Control input constraint: \(\mathcal{U} = [-0.75, 1]\)
		\item State constraints: \(x_1 \in [-0.1, 1.3], \quad x_2 \in [-0.4, \infty)\)
	\end{itemize}
For the lifted NMPC \eqref{eq:liftedmpc}, the subdivision factor \(N'\) for the FSFH approximation is set to 10.
The optimization is carried out using \texttt{fmincon}, a nonlinear constrained optimization solver from MATLAB's Optimization Toolbox.
The comparison between the results of the conventional NMPC and the lifted NMPC controllers is depicted in \cref{fig:combined_vdp}, where the lifted NMPC demonstrates more efficient steering.

\subsection{Inverted Pendulum on a Cart}
Consider the problem of swing-up control for an inverted pendulum on a cart, as illustrated in \cref{fig:cartpendulum}.
The system's dynamics are described by the following set of nonlinear differential equations:
	\begin{equation} \label{eq:cartpendulum}
		\dot{x}
			=
			\begin{bmatrix}
				\dot{x}_1 \\
				\dot{x}_2 \\
				\dot{x}_3 \\
				\dot{x}_4
			\end{bmatrix}
			=
			\begin{bmatrix}
				x_3 \\
				x_4 \\
				\frac{- m_{p} l x_{4}^{2} \sin(x_{2}) + m_{p} g \sin(x_{2}) \cos(x_{2}) + u}{m_{c} + m_{p} \sin^2(x_{2})} \\
				\frac{- m_{p} l x_{4}^{2} \sin(x_{2}) \cos(x_{2}) + (m_{c} + m_{p}) g \sin(x_{2}) + u \cos(x_{2})}{l (m_{c} + m_{p} \sin^2(x_{2}))}
			\end{bmatrix},
	\end{equation}
where \(x_1\) represents the horizontal position of the cart, \(x_2\) is the angular displacement of the pendulum from the upright position, \(x_3\) is the translational velocity of the cart, and \(x_4\) is the angular velocity of the pendulum.
The control input \(u\) is the force applied to the cart.
The system parameters and their values are summarized in \cref{tab:inverted_pendulum_parameters}.
The objective is to swing the pendulum up to the upright position, starting from the initial state \(\left(x_1(0), x_2(0), x_3(0), x_4(0)\right) = (0, \pi, 0, 0)\).

In this example, we first consider single-rate control and compare the performance of the conventional NMPC with that of the lifted NMPC, as in the previous example. Subsequently, we introduce multirate control (\(M\geq 2\) in \eqref{eq:multirateinput}), and demonstrate that it can effectively control the system even with a slow sampling period, where single-rate control fails.
In all the cases, the optimization problems \eqref{eq:fin_mpc} and \eqref{eq:liftedmpc} are solved using OpEn \cite{sopasakis2020open}, an optimizer that uses the PANOC \cite{stella2017simple}, a fast and accurate embedded numerical solver well-suited for real-world applications.
This framework has been adopted in various studies, including \cite{cataffo2022nonlinear, lindqvist2020nonlinear, lindqvist2024tree, lindqvist2021scalable, mansouri2020subterranean, nag2022flock, nag2024distributed}.
OpEn supports two approaches for handling state constraints: the augmented Lagrangian method (ALM) and the penalty method.
The ALM introduces Lagrange multipliers and iteratively enforces constraint satisfaction more accurately, while the penalty method embeds constraint violations directly into the cost function, trading off accuracy for faster computation.
In our simulations, we compare both methods in terms of computation time and discuss the feasibility of real-time implementation with the proposed lifted NMPC framework.

	\begin{minipage}[b]{0.5\linewidth}
		\centering
		\begin{tabular}{|c|c|c|c|}
			\hline
			\textbf{Parameter} & \textbf{Description} & \textbf{Value} & \textbf{Units} \\ \hline
			\(g\)  & Acceleration due to gravity & 9.8 & m/s$^2$ \\ \hline
			\(l\) & Length of the pendulum & 1.0 & m \\ \hline
			\(m_c\)  & Mass of the cart & 1.0 & kg \\ \hline
			\(m_p\)  & Mass of the pendulum & 0.2 & kg \\ \hline
		\end{tabular}
		\captionof{table}{Inverted pendulum system's parameters.}
		\label{tab:inverted_pendulum_parameters}
	\end{minipage}
	\hfill
	\begin{minipage}[b]{0.45\linewidth}
		\centering
		\includegraphics[width=\linewidth]{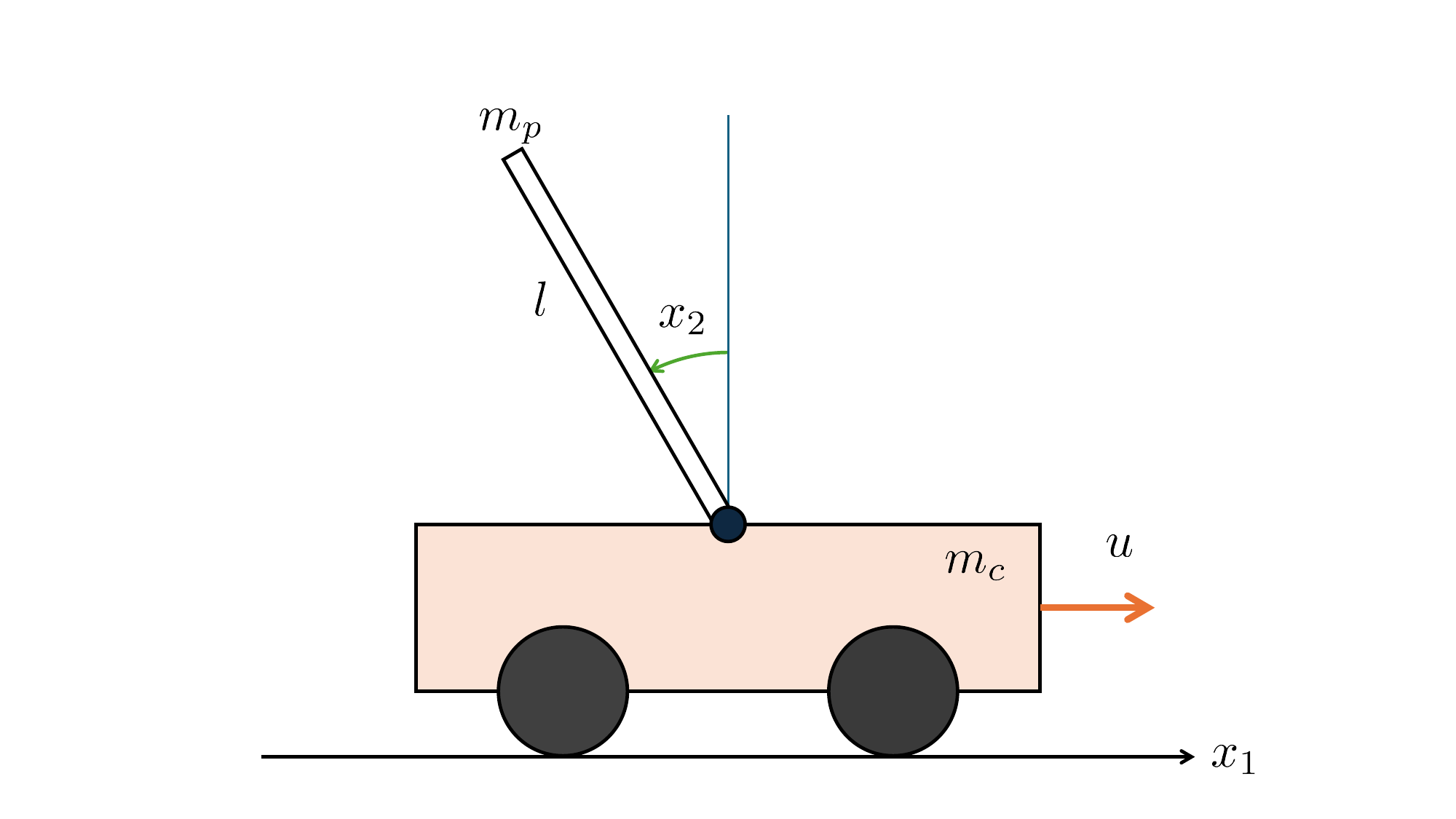}
		\captionof{figure}{Inverted pendulum on a cart model.}
		\label{fig:cartpendulum}
	\end{minipage}

\subsubsection{Single-Rate Control}
As in the previous example, we first compare the performance of the conventional NMPC and the lifted NMPC when the control input is set constant between the sampling instants as \(u(t) = u_{\rm d}[k]\) and \(u(t) = v[k]\) for \(kT\leq t < (k+1)T\), \(k=0,1,\dots\), obtained by solving \eqref{eq:fin_mpc} and \eqref{eq:liftedmpc}, respectively.

The optimization variables are set as follows:
	\begin{itemize}
		\item Sampling period: \(T=0.05\) (s)
		\item Prediction horizon: \(N=20\)
		\item Cost function weights:
			\(
				Q = \diag(2.5,10.0,0.01,0.01),
				\quad
				R = 0.1,
				\quad
				Q' = \diag(3.0,10.0,0.02,0.02)
			\)
		\item Control input constraint: \(\mathcal{U} = [-15, 15]\)
		\item State constraints: \(x_1 \in [-2.5,5.0], \quad x_3 \in [-7.5,7.5]\)
	\end{itemize}
The subdivision factor \(N'\) for the FSFH approximation is set to 10 for the lifted NMPC \eqref{eq:liftedmpc}.

\begin{figure}[tb]
		\centering
		\begin{subfigure}{0.48\linewidth}
			\centering
			\includegraphics[width=\linewidth]{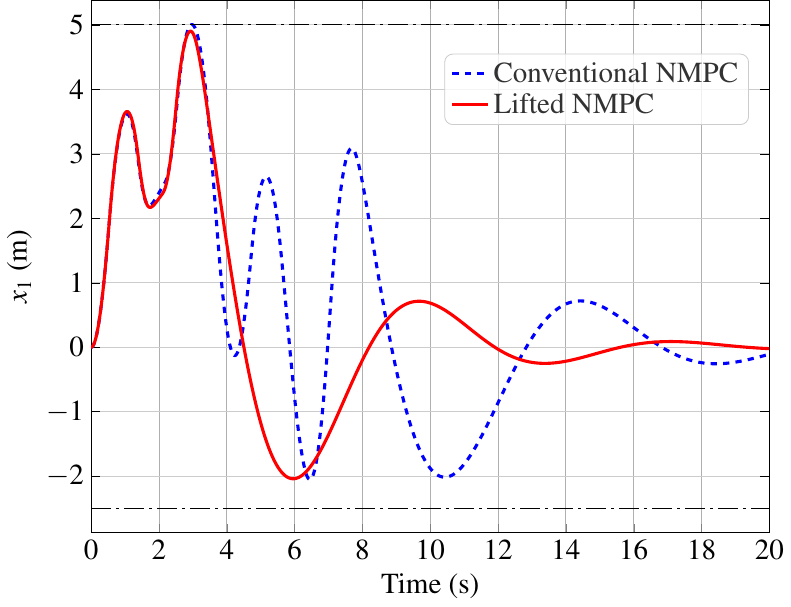}%
		\end{subfigure}
		\hfill
		\begin{subfigure}{0.48\linewidth}
			\centering
			\includegraphics[width=\linewidth]{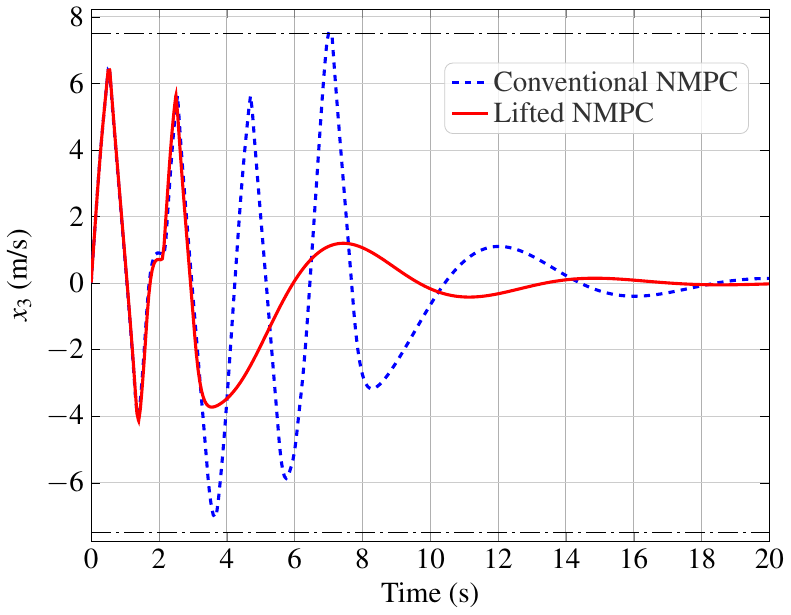}%
		\end{subfigure}

		\vspace{0.5cm} 

		\begin{subfigure}{0.48\linewidth}
			\centering
			\includegraphics[width=\linewidth]{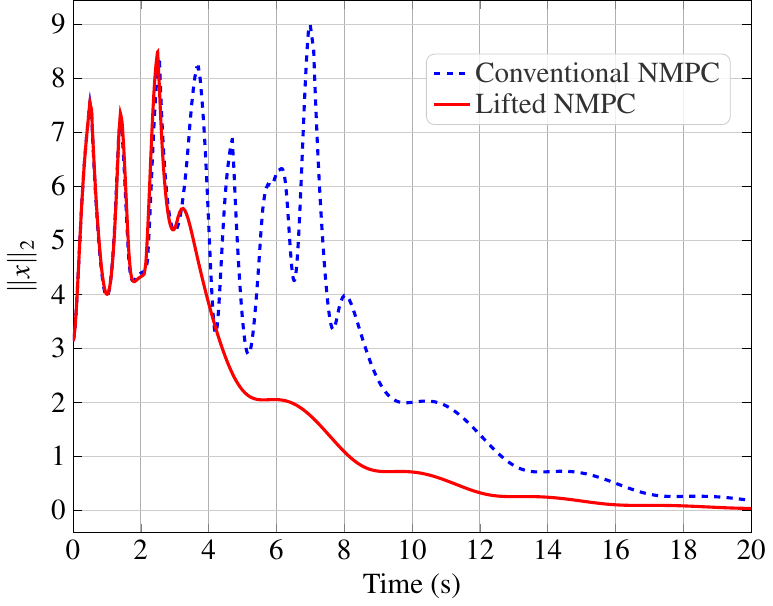}%
		\end{subfigure}
		\hfill
		\begin{subfigure}{0.48\linewidth}
			\centering
			\includegraphics[width=\linewidth]{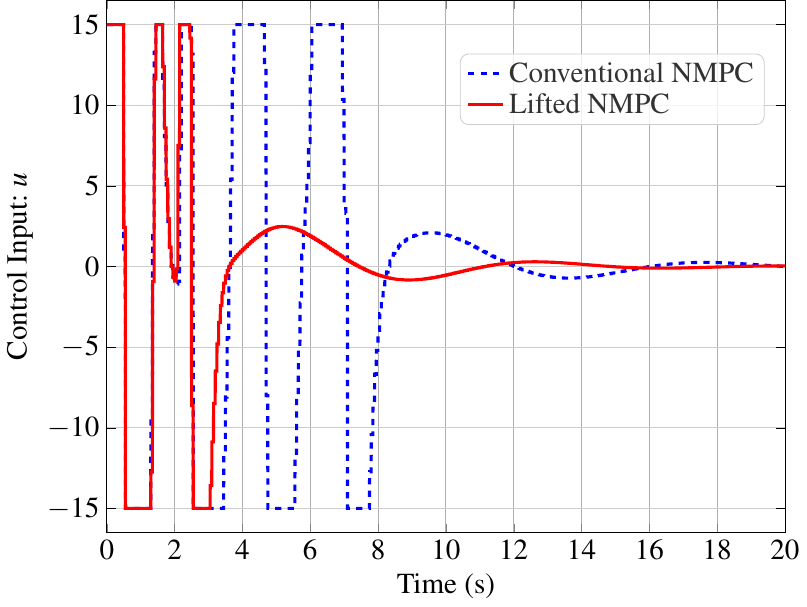}%
		\end{subfigure}
		\caption{State and input trajectories of the inverted pendulum on a cart using conventional NMPC (blue dashed) and lifted NMPC (red solid). Time-series plots of the cart position \(x_1\) and velocity \(x_3\) are shown in the top-left and top-right plots, respectively, with black dashed-dotted lines indicating state constraint boundaries. The bottom-left plot shows the time series of the Euclidean norm of all four states, whereas the bottom-right plot shows the control input \(u\).}
		\label{fig:combined_ipc}
	\end{figure}

The state trajectories (\(x_1\) and \(x_3\)), the Euclidean norm \(\|x(t)\|\) of the state trajectories, and the control inputs for both the conventional and lifted NMPC approaches are compared in \cref{fig:combined_ipc}.
Here, the Euclidean norm is given by \(\| x(t)\| = \sqrt{\sum_{i=1}^4x_{i}(t)^2}\).
As previously discussed, we use both ALM and the penalty method to handle state constraints and compare the results.
Since both methods yielded identical trajectories, we present only the results for ALM.
The corresponding solve times for each method are presented in \cref{fig:solve_time_ipc_alm,fig:solve_time_ipc_pm}, respectively.
While the control performance is unaffected, the penalty method significantly reduces computation time.
In the case of ALM, occasional spikes in solve time exceed the 50 ms control interval, likely due to numerical issues.
However, these spikes occur infrequently and have negligible impact on control performance.
It is also possible to impose a maximum solve time, allowing the solver to terminate early while still providing a solution of acceptable quality for real-time application.
This approach is used in the next section.

Simulation videos are also available at {\sf\url{https://youtu.be/lPSm05bjKLk}} and {\sf\url{https://youtu.be/DjkAUKNTPcU}}.
Both the figures and videos demonstrate that the lifted NMPC approach achieves superior performance.
As shown in \cref{tab:inverted_pendulum_performance}, the lifted NMPC consistently achieves faster settling times across all state variables.
Moreover, the root mean square (RMS) values of the state norm \(\|x(t)\|\) and the control input, which are directly related to the energy expenditure, are significantly lower.
However, this improvement comes at the cost of increased average solve time per control step.
This overhead arises from solving a higher dimensional optimization problem in the lifted formulation, leading to longer solve times compared to conventional NMPC under both ALM and the penalty method.

	\begin{minipage}[t]{0.48\linewidth}
			\centering
			\includegraphics[width=\linewidth]{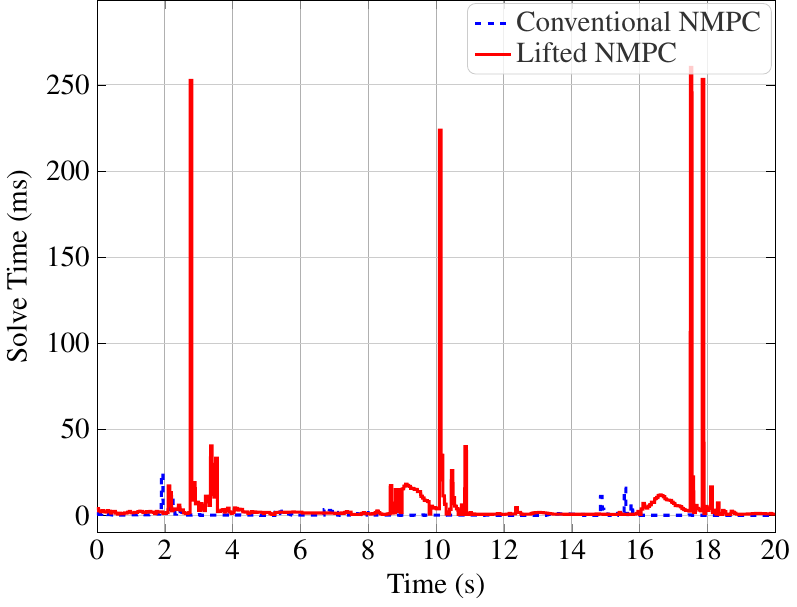}%
			\captionof{figure}{Solve time required in a control loop during the experimentation for both the conventional and lifted NMPCs, where ALM is used to handle state constraints.}
			\label{fig:solve_time_ipc_alm}
	\end{minipage}
	\hfill
	\begin{minipage}[t]{0.48\linewidth}
			\centering
			\includegraphics[width=\linewidth]{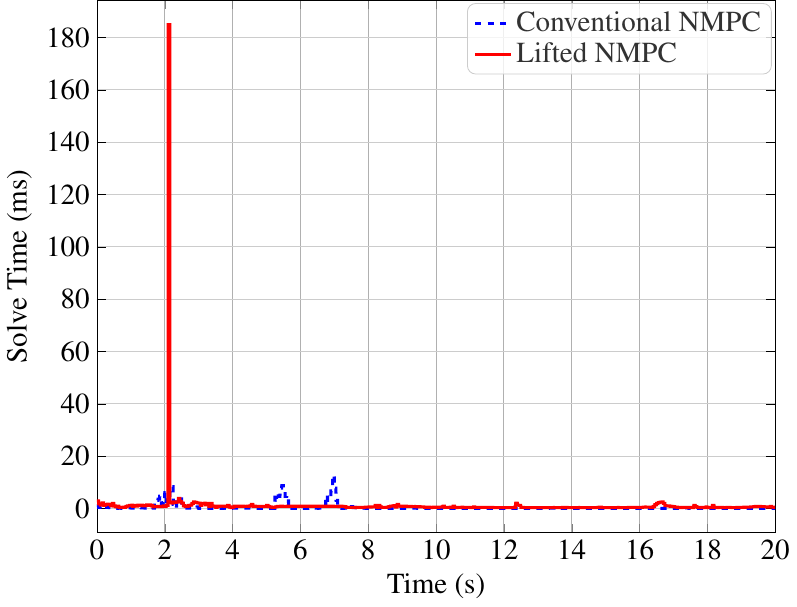}%
			\captionof{figure}{Solve time required in a control loop during the experimentation for both the conventional and lifted NMPCs, where the penalty method is used to handle state constraints.}
			\label{fig:solve_time_ipc_pm}
	\end{minipage}

%

	\begin{table}[htb]
		\centering
		\begin{tabular}{|l|c|c|c|}
		\hline
		\textbf{Metric} & \textbf{Conventional NMPC} & \textbf{Lifted NMPC} & \textbf{Units} \\ \hline
			\(x_1\) settling time  & 23.821 & 18.299 & m \\ \hline
			\(x_2\) settling time & 14.447 & 9.815 & rad \\ \hline
			\(x_3\) settling time  & 24.699 & 18.860 & m/s \\ \hline
			\(x_4\) settling time  & 12.668 & 8.144 & rad/s \\ \hline
                RMS of \(\|x \|\)  & 3.766 & 2.677 & --- \\ \hline
                RMS of control input  & 8.538 & 5.417 & N \\ \hline
                Average solve time (ALM)  & 0.475 & 3.527 & ms \\ \hline
                Average solve time (penalty method)  & 0.351 & 0.980 & ms \\ \hline
		\end{tabular}
		\caption{Comparison of performance metrics between conventional and lifted NMPC for the inverted pendulum on a cart over 20 s. RMS stands for root mean square.}
		\label{tab:inverted_pendulum_performance}
	\end{table}

\subsubsection{Multirate Control}
Since the lifted NMPC is designed to minimize a cost using a sequence of control input functions \(\{u[k](\theta)\}_{k=0}^{N-1}\), it is natural to consider multirate control \eqref{eq:multirateinput} where the control frequency is higher than the sampling frequency, which more clearly demonstrates the advantage of the lifted NMPC.

The optimization variables and parameters are set as follows:
	\begin{itemize}
		\item Prediction horizon period: \(1.0\) (s)
		\item Cost function weights:
			\(
				Q = \diag(2.5,10.0,0.01,0.01),
				\quad
				R = 0.1,
				\quad
				Q' = \diag(3.0,30.0,0.1,0.02)
			\)
		\item Control input constraint: \(\mathcal{U} = [-20, 20]\)
		\item State constraints: \(x_1 \in [-7,7], \quad x_3 \in [-15,15]\)
	\end{itemize}
and the subdivision factor \(N'\) for the FSFH approximation is set to 10 for both single-rate lifted NMPC and multirate lifted NMPC.

We then consider three different sampling periods, \(T=0.1, 0.25,\) and 0.5 s.
For the multirate lifted NMPC, the control period is fixed at 0.05 s, resulting in upsampling rates \(M=2, 5,\) and 10, respectively\footnote{For implementation, the subdivision factor \(N'\) should be set as an integer multiple of the upsampling factor \(M\).}.
In this experimentation, we use ALM to handle the state constraints, setting the maximum solver duration to 45 ms (i.e., shorter than the control period) to ensure real-time feasibility.

\begin{figure}[tb]
	\centering
	\includegraphics[width=0.6\linewidth]{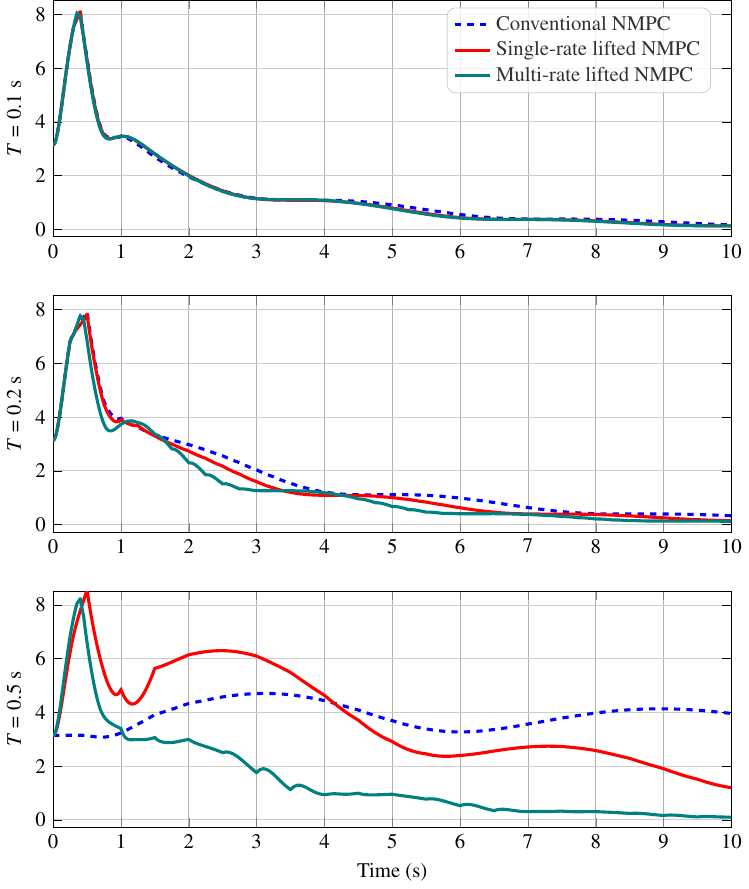}%
 	\caption{Euclidean norm of the state over time for conventional NMPC (blue dashed), lifted NMPC (red solid), and multirate lifted NMPC (green solid) with sampling periods \(T=0.1,0.25,0.5\) s.}
	\label{fig:2norm}
\end{figure}

\begin{figure}[htb]
	\centering
	\includegraphics[width=0.6\linewidth]{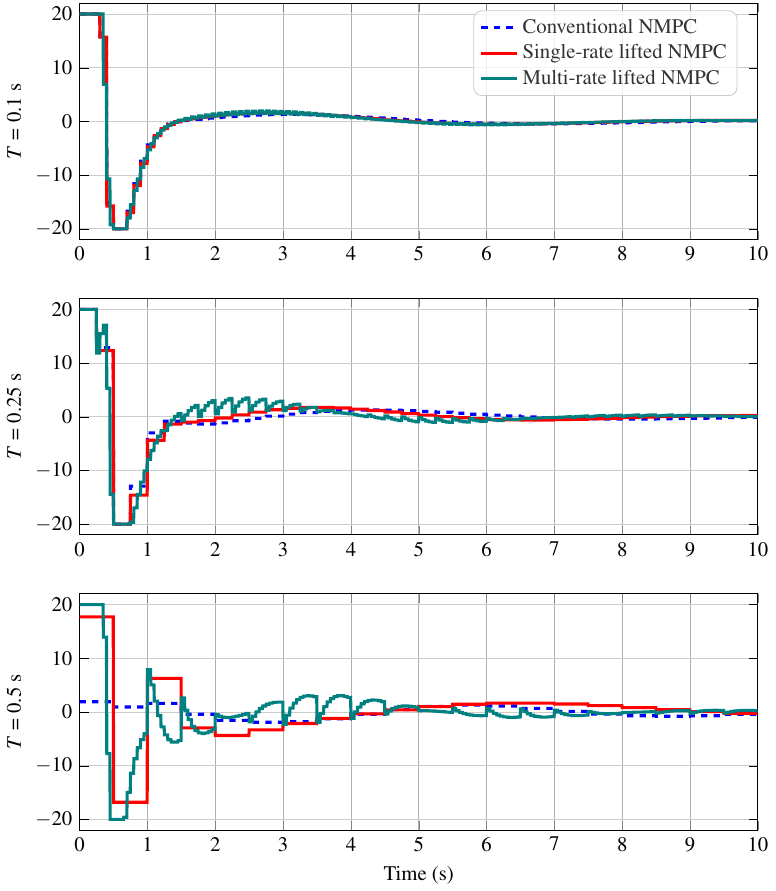}%
	\caption{Control input of conventional NMPC (blue dashed), lifted NMPC (red solid), and multirate lifted NMPC (green solid) with sampling periods \(T=0.1,0.25,0.5\) s.}
	\label{fig:mrinput}
\end{figure}

\begin{figure}[htb]
    \centering
    \includegraphics[width=0.6\linewidth]{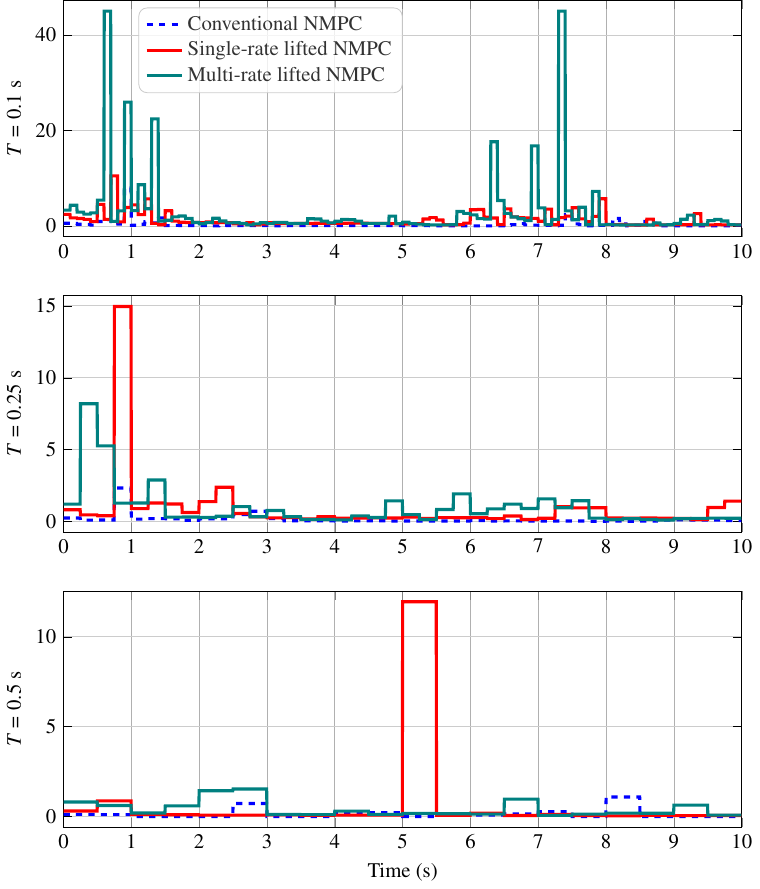}%
    \caption{Solve time required in a control loop over the experimentation of conventional NMPC (blue dashed), lifted NMPC (red solid), and multirate implemented lifted NMPC (solid green) with sampling periods \(T=0.1,0.25,0.5\) s.}
    \label{fig:mrsolvetime}
\end{figure}

The plots of the Euclidean norm of all states at each time for the conventional NMPC, the single-rate lifted NMPC, and the multirate lifted NMPC are shown in \cref{fig:2norm}, with \(T=0.1, 0.25,\) and 0.5 s.
The control inputs and the solve time required per control loop for each case are shown in \cref{fig:mrinput,fig:mrsolvetime}, respectively.
The simulation video is available at {\sf\url{https://youtu.be/sVLA8YHhU0E}}.
As depicted in the plots and \cref{tab:inverted_pendulum_performance_multirate}, the multirate lifted NMPC outperforms both the single-rate lifted NMPC and the conventional NMPC, particularly at longer sampling periods.
Specifically, for \(T=0.5\) s, the conventional NMPC fails to achieve swing-up, while the single-rate lifted NMPC---although able to swing the pendulum up---shows degraded performance compared to the multirate approach, which successfully achieves swing-up and stabilization with better overall performance.

\begin{table}[htb]
    \centering
    \begin{tabular}{|c|l|c|c|c|c|}
        \hline
        \textbf{Sampling Period} & \textbf{Metric} & \textbf{Conventional NMPC} & \textbf{Single-rate Lifted NMPC} & \textbf{Multirate Lifted NMPC} & \textbf{Units} \\ \hline
        \multirow{3}{*}{T = 0.1 s}
            & RMS of \(\|x\|\) & 2.036 & 2.032 & 2.025 & --- \\ \cline{2-6}
            & RMS of control input & 5.484 & 5.537 & 5.508 & N\\ \cline{2-6}
            & Average solve time & 0.354 & 1.243 & 3.745 & ms \\ \hline
        \multirow{3}{*}{T = 0.25 s} 
            & RMS of state norm & 2.397 & 2.285 & 2.150 & --- \\ \cline{2-6}
            & RMS of control input & 5.385 & 5.496 & 5.664 & N \\ \cline{2-6}
            & Average solve time & 0.161 & 0.897 & 0.964 & ms \\ \hline
        \multirow{3}{*}{T = 0.5 s} 
            & RMS of state norm & 3.925 & 4.193 & 2.210 & ---\\ \cline{2-6}
            & RMS of control input & 1.116 & 5.892 & 5.517 & N \\ \cline{2-6}
            & Average solve time & 0.176 & 0.749 & 0.431 & ms\\ \hline
    \end{tabular}
    \caption{Comparison of the root mean square (RMS) state norms, control input, and solve times for conventional, single-rate lifted, and multirate lifted NMPC over 10 s with varying sampling periods.}
    \label{tab:inverted_pendulum_performance_multirate}
\end{table}

Moreover, although the solve times required for the lifted NMPC are comparatively higher than for the conventional one, they are still fast enough for real-time online NMPC implementation, enabling more accurate and robust trajectory tracking in complex environments.
This trade-off between computational complexity and control performance is justified by the enhanced ability of the lifted approach to handle nonlinearities more effectively, ultimately leading to improved system behavior and safety in practical scenarios.

%% file: TeX/Text/Conclusion.tex
In this paper, we introduced a novel nonlinear model predictive control (NMPC) framework that leverages nonlinear lifting techniques to enhance the control of nonlinear systems.
Using two illustrative examples—the Van der Pol oscillator and the inverted pendulum on a cart—we demonstrated that the lifted NMPC outperforms the conventional NMPC, particularly in reducing settling time, improving control efficiency, and handling intersample behavior.

Additionally, we extended the framework to a multirate lifted NMPC, where the control frequency exceeds the sampling frequency.
Simulation results showed that the multirate approach achieves superior performance, particularly under slow sampling rates, where conventional and single-rate lifted NMPCs struggle to maintain control.

By incorporating an FSFH approximation for the dynamics and using standard numerical methods such as Runge--Kutta for state trajectory estimation and Simpson's rule for cost integration, we achieved a computationally efficient implementation of the lifting-based NMPC.

These results underscore the potential of the lifted NMPC framework, including its multirate extension, for practical applications requiring optimal control of nonlinear and highly dynamic systems.
Future research will explore the application of this approach to more complex nonlinear systems and its real-time implementation in embedded control platforms.